\documentclass[11pt,twoside]{article}

\usepackage{galev06}
\usepackage{epsf}
\usepackage{psfig}
\usepackage{lscape}

\newcommand{\Msolar}{\mbox{$M_{\odot}\,$}}

\def\gs{\mathrel{\raise0.35ex\hbox{$\scriptstyle >$}\kern-0.6em \lower0.40ex\hbox{{$\scriptstyle \sim$}}}}
\def\ls{\mathrel{\raise0.35ex\hbox{$\scriptstyle <$}\kern-0.6em \lower0.40ex\hbox{{$\scriptstyle \sim$}}}}

\begin{document}
\title{Dense Molecular Gas in Extreme Starburst Galaxies -- What will we learn from Herschel?}   
\author{T.\ R.\ Greve}   
\affil{California Institute of Technology, Pasadena, CA 91125, USA}    
\author{P.\ P.\ Papadopoulos}   
\affil{Institut f\"{u}r Astronomie, ETH, Z\"{u}rich 8093, Switzerland}    
\author{Y.\ Gao}   
\affil{Purple Mountain Observatory, Chinese Academy of Sciences, Nanjing 210008, People's Republic  of China}    
\author{S.\ J.\ E.\ Radford}   
\affil{California Institute of Technology, Pasadena, CA 91125, USA}    

\begin{abstract} 
Ultra Luminous Infra-Red Galaxies (ULIRGs) 
 -- gas-rich mergers harboring the most extreme 
star-forming conditions encountered in the local Universe -- 
are thought to re-enact the galaxy formation processes we are 
only barely able to glimpse in the distant Universe.
Combining new single-dish molecular line observations of
$^{12}$CO, $^{13}$CO, HCO$^+$, HCN, and CS towards the two ULIRGs Arp\,220 and NGC\,6240
with existing data in the literature, we have compiled 
the most extensive molecular line data-sets to date of such galaxies.
The data allow us to put strong constraints on the properties of the dense star forming gas 
in these two systems, and compare the relative usefulness of CS, HCN and HCO$^+$ 
as tracers of dense gas. In addition, we have build molecular line templates based on our observations,
and demonstrate that Herschel/HI-FI will be able to detect the high-$J$ 
transitions of most of the above molecules in a large sample of ULIRGs out to $z\ls 0.5$,
assuming Arp\,220 and NGC\,6240 are representative of the ULIRG population at these redshifts.
\end{abstract}


\section{Introduction}   
The coming years will witness the emergence of a  
new generation of ground-based and space-born mm/sub-mm telescopes
with sensitivities and frequency coverage that will
allow us to study the interstellar medium (ISM) in 
high-$z$ proto-galaxies in unprecedented detail.
Unfortunately, our expectations of how such objects will look in mm/sub-mm line emission
are ill-informed due to the dearth of similar data locally.
Virtually all studies of local (U)LIRGs to date have been of
low-$J$ CO transitions or of single HCN/HCO$^+$ lines, and as a result
we know very little about the properties of their dense, star-forming gas phase.
While pioneering multi-line studies of low-intensity starbursts taught us the
value of having many independent dense gas tracers (Nguyen-Q-Rieu et al.\ 1992; Jackson et al.\ 1995; 
Paglione, Jackson \& Ishizuki 1997), the relatively benign star formation
rates of such galaxies suggest they may not be optimal guides 
for the state of the ISM in (U)LIRGs and high-$z$ starbursts.
In the local Universe, Arp\,220 and NGC\,6240 are ideally suited for 
studies of the dense gas in extreme starburst galaxies 
as they have the most well-sampled FIR/sub-mm spectral energy 
distributions (SEDs) of their category (Dopita et al.\ 2005) and a large number of
rotational lines of CO -- and some of HCN, CS and HCO$^+$ -- have already been observed.
With this in mind, we used the 15-m James Clerk Maxwell Telescope 
and the IRAM 30-m Telescope, to observe
HCN, CS, and HCO$^+$ molecular line emission towards these two sources (Greve et al.\ 2006).

\section{The dense molecular ISM in Arp\,220 and NGC\,6240}
The final tally of high-density tracers observed is CS 2--1, 3--2, 5--4, 7--6 (Arp\,220 only), 
HCN 1--0, 3--2, 4--3, and HCO$^+$ 1--0, 3--2, 4--3. 
Employing a Large Velocity Gradient (LVG) to model the line ratios from each molecular species,
we find that while HCO$^+$, HCN, and CS all trace what in broad terms can be characterized as dense gas 
($n(\mbox{H}_2) \gs 10^4\,\mbox{cm}^{-3}$), the individual solutions point toward significantly
different densities. {\it In fact, we find that the CS emission emerges from gas with a density
10$\times$ higher than what is derived from the HCN line ratios, and 100$\times$ denser 
than the HCO$^+$ emitting gas.} This is consistent with Galactic surveys of CS lines, which have shown them to be
unmistakable markers of very dense, high-mass star forming cores (Plume et al.\ 1997; Shirley et al.\ 2003), while
HCO$^+$ with its smaller critical density is expected to trace
less dense gas than both CS and HCN. 
Thus, our findings strongly suggest that the recent claims that HCO$^+$ is a better tracer of the dense
gas in (U)LIRGs (Graci\'{a}-Carpio et al.\ 2006) does not hold in Arp\,220 and NGC\,6240.
The temperature constraints provided
by the three LVG solutions are remarkably similar ($T_k\simeq 50-70$\,K).  
This may indicate that HCO$^+$, HCN, and CS all trace the star forming part of the molecular gas reservoir,
especially since the $T_k$-values are rather high, which could be due to the proximity of the gas phase probed
to the star formation sites in both galaxies.

Assuming the dense gas resides in virialized and optically thick (in HCN 1--0) clouds, 
the associated total gas mass is given by 
\begin{equation}
M_{dense}(\mbox{H}_2) = \alpha_{\mbox{\tiny{HCN}}} L'_{\mbox{\tiny{HCN}}}, 
\end{equation}
where $L'_{\mbox{\tiny{HCN}}}$ is the HCN 1--0 line luminosity, and 
$\alpha_{\mbox{\tiny{HCN}}}\simeq 2.1 \sqrt{n(\mbox{H}_2)}/T_{b,\mbox{\tiny{HCN(1-0)}}}$
is the HCN-to-H$_2$ conversion factor (e.g.\ Radford, Solomon \& Downes 1991) and is estimated
from the best-fit LVG solution to the HCN line ratios.
In a similar fashion, we can derive conversion factors and dense gas masses appropriate for HCO$^+$ and CS.
Combining the dense gas mass estimates obtained in this way, we find 
$M_{dense} \simeq (0.6-2.5)\times 10^{10}\,\Msolar$ and 
$M_{dense} \simeq (0.8-2.4)\times 10^{10}\,\Msolar$ for Arp\,220 and
NGC\,6240, respectively.
Thus in both systems we find a remarkably narrow range in dense gas
mass estimates and, given our strong constraints on the dense gas properties, 
they are likely to be amongst the most accurate ever obtained for any (U)LIRG.
While consistent with dynamical estimates of the total
mass, such large dense gas masses strongly suggest that the bulk
of the total gas reservoir in Arp\,220 and NGC\,6240 is in the dense phase --
consistent with the notion that most of their IR luminosities are powered by star formation.

\section{Probing the dense gas at high redshifts with Herschel}	
Evidence from Galactic observations points to a common
gas phase being responsible for the HCN and high-$J$ CO line emission.
This is indeed expected, as such a gas phase is intimately linked with 
ongoing star formation, so its warm and dense conditions are able
to excite HCN, CS, HCO$^+$ as well as the high-$J$ CO transitions.
If this correlation between the high-$J$ CO and HCN, CS, HCO$^+$ line emission 
is verified for (U)LIRGs in general, future ground-based large aperture telescopes
(ALMA, CCAT, LMT) will be able to capitalize from it, as it 
offers a more accessible observational window to the dense
star forming molecular gas in galaxies, with the latter lines being much easier to observe
than the high-$J$ CO lines whose emission at $\gs 460$\,GHz is severely attenuated
by the Earth's atmosphere. 

\begin{figure}[h]
\plotone{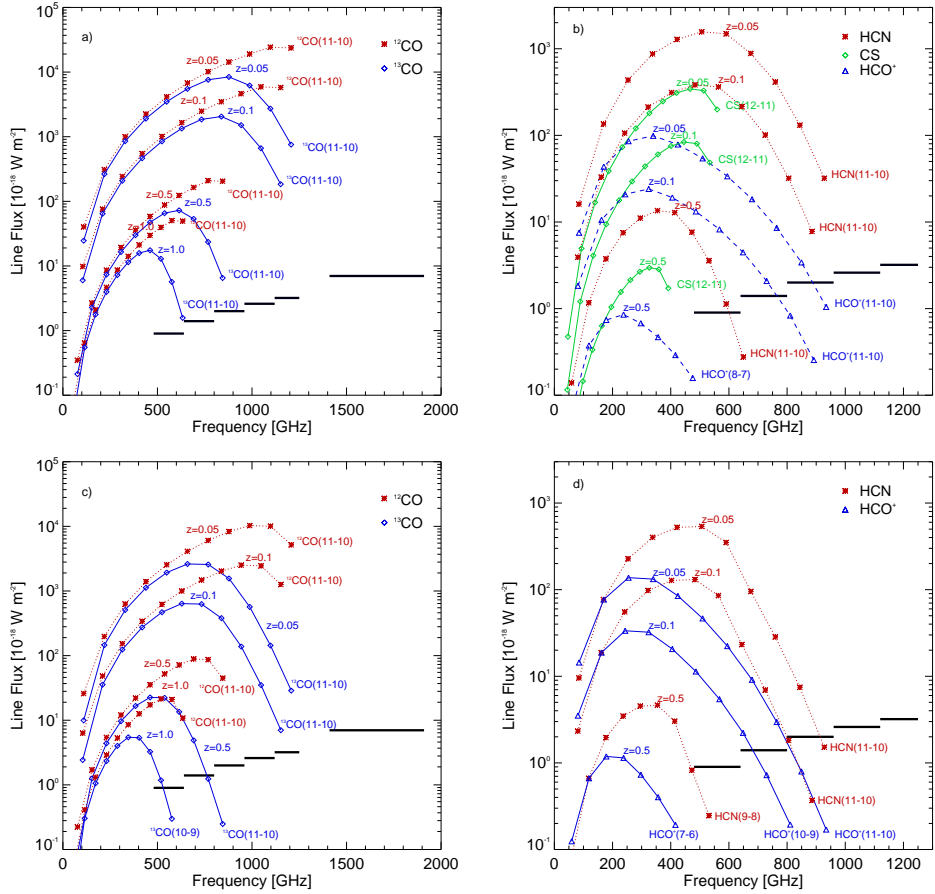}
\caption{{\bf Top panels:} integrated line flux strengths of the full $^{12}$CO/$^{13}$CO
(left panels) and HCN/CS/HCO$^+$ (right panels) rotational $J$-ladders, as predicted by the dense gas template 
of Arp\,220. The rotational transitions start at $J=1-0$ (on the left hand side of each plot),
and are in most cases plotted consecutively up to $J=11-10$ (on the right hand side of each plot).
The line fluxes are shown for Arp\,220-like systems at redshifts $z=0.05, 0.1, 0.5$, and $1.0$.
{\bf Bottom panels:} Same as above, except the constraints on the dense gas in 
NGC\,6240 have been used as a template. In all panels, the horizontal bars indicate the predicted
$5$-$\sigma$ sensitivity limits of the six HI-FI bands after 1\,hr of integration. 
}
\label{figure:predictions}
\end{figure}

The HI-FI instrument (de Graauw et al.\ 1998) onboard Herschel 
should allow us to investigate whether the aforementioned concomitance of molecular line
emission is widespread amongst starburst and (U)LIRGs.
We have therefore used our constraints on the dense gas properties
in Arp\,220 and NGC\,6240 to predict the line fluxes of these
transitions and the extent HI-FI  
will be able to detect these lines in (U)LIRGs at various redshifts (Fig.\ \ref{figure:predictions}). 
For (U)LIRGs in the redshift range $0\ls z \ls 0.5$, Herschel/HI-FI will 
complement ground-based (sub)mm facilities for which currently only the low-$J$
lines are accessible and thereby virtually complete the full rotational $J$-ladder
for a number of important molecules in such systems.
Not only will this allow for a full inventory of the molecular
ISM in luminous starbursts and (U)LIRGs, but it will also provide us
with a valuable data-base of low-$z$ templates with which to interpret
the high-$J$ detections in extremely distant galaxies ($z\gs 2$) that
ALMA will produce. Characterizing these transitions relative to the lower
ones for the local extreme starburst population  -- similar to the need of
understanding local dust SEDs before interpreting the single-frequency detections
of dust continuum in high-$z$ galaxies -- will prove to be extremely
valuable before embarking on the high-$J$, high-$z$ ventures with ALMA.

\acknowledgements 
We are grateful to the telescope operators on the JCMT and IRAM 30-m Telescope for
their help with taking the data. We thank the organizers of the conference for putting
together an excellent meeting.


\end{document}